# Intracluster Globular Clusters

Michael J. West,[1] Patrick Côté,[2] Christine Jones,[3] William Forman[3] and Ronald O. Marzke[2]

## ABSTRACT

Globular cluster populations of supergiant elliptical galaxies are known to vary widely, from extremely populous systems like that of UGC 9799, the centrally dominant galaxy in Abell 2052, to globular-cluster-poor galaxies such as NGC 5629 in Abell 2666. Here we propose that these variations point strongly to the existence of a population of globular clusters that are not bound to individual galaxies, but rather move freely thoughout the cores of clusters of galaxies. Such intracluster globular clusters may have originated as tidally stripped debris from galaxy interactions and mergers, or alternatively they may have formed *in situ* in some scenarios of globular cluster formation.

*Subject headings:* galaxies: clusters: general – galaxies: elliptical and lenticular, cD — galaxies: star clusters – X-rays: galaxies

---

[1] Department of Astronomy & Physics, Saint Mary's University, Halifax, NS B3H 3C3, Canada

[2] Dominion Astrophysical Observatory, Herzberg Institute of Astrophysics, National Research Council, 5071 West Saanich Road, Victoria, BC V8X 4M6, Canada

[2] Harvard-Smithsonian Center for Astrophysics, 60 Garden Street, Cambridge, MA 02138, USA

## 1. Introduction

One of the outstanding problems in globular cluster (hereafter GC) research is the origin of the extremely populous GC systems around many, though not all, supergiant elliptical galaxies. A useful measure of a galaxy's GC population is its *specific globular cluster frequency*, $S_N$, defined as

$$S_N = N_t \times 10^{0.4(M_V+15)} \qquad (1)$$

where $N_t$ is the total number of globular clusters and $M_V$ is the absolute visual magnitude of the galaxy (Harris & van den Bergh 1981). To date, specific frequencies have been measured for 14 brightest cluster galaxies (hereafter BCGs); these data are summarized in Table 1.[4] A wide range of specific frequencies is seen, from $S_N \simeq 4$ to $S_N \simeq 20$. The lower limit of $S_N \simeq 4-5$ for BCGs is similar to the typical value found for most normal elliptical galaxies in dense environments (e.g., Harris 1991). This suggests that low-$S_N$ BCGs have normal GC populations characteristic of elliptical galaxies in general, whereas high-$S_N$ BCGs have anomalously rich GC populations. The prototypical high-$S_N$ galaxy is M87 in the Virgo cluster, which possesses more than 15,000 GCs, roughly three times as many GCs per unit luminosity as any other Virgo elliptical, and nearly an order of magnitude more than typical field galaxies.

The reason for the dichotomy between high-$S_N$ and low-$S_N$ BCGs is unknown at present. No obvious correlation exists between $S_N$ and any intrinsic property of either the BCG or the galaxy cluster in which it resides (Harris et al. 1995). As Figure 1 shows, for example, $S_N$ appears to be independent of cluster X-ray temperature, which is proportional to the total cluster mass. Furthermore, no galaxy cluster has been found to have more than one high-$S_N$ galaxy.

In this *Letter* we propose that the observed $S_N$ variations of BCGs point strongly to the existence of a population of intracluster globular clusters (hereafter IGCs) that are not bound to individual galaxies, but rather move freely throughout the potential wells of galaxy clusters. We note that this idea is not a new one; this possibility has previously been considered, and subsequently rejected, by a number of authors over the years (e.g., Fabian, Nulsen & Canizares 1984; van den Bergh 1984; Harris 1986; Muzzio 1987). However, as we shall show below, a simple model based on the IGC hypothesis can account quite successfully for the high-$S_N$ phenomenon.

---

[4]NGC 4874 and NGC 4889 in the Coma cluster (Abell 1656) have been included since both are supergiant elliptical galaxies which are clearly the dominant cluster members. For this same reason both M87 (NGC 4486) and M49 (NGC 4472) in the Virgo cluster have also been included.

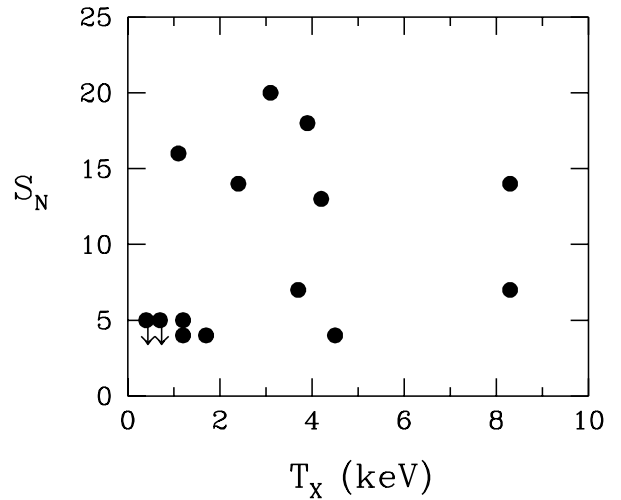

Figure 1. $S_N$ versus galaxy cluster X-ray temperature $T_X$. $S_N$ values for NGC 5629 and NGC 5424 are upper limits.

## 2. Intracluster Globular Clusters: A Model

We propose that the apparent excess of GCs associated with high-$S_N$ galaxies originates from a population of intracluster globular clusters. We construct a quantitative model based on the following specific assumptions:

- There exists a population of intracluster globular clusters in all galaxy clusters. The total number of IGCs depends on the cluster mass

$$N_{IGC} \propto M_{cl} \qquad (2)$$

with richer galaxy clusters having more IGCs than poorer clusters.

- The distribution of IGCs follows the cluster mass distribution. A convenient functional form which satisfactorily describes the projected distribution of galaxies and the dark matter component indicated by gravitational lens studies is the

King (1962) approximation of a bounded isothermal sphere

$$\Sigma(r) = \frac{\Sigma_o}{(1 + r^2/r_c^2)} \quad (3)$$

where $\Sigma(r)$ is the projected mass density at a distance $r$ from the galaxy cluster center, $\Sigma_o$ is the central density, and $r_c$ is the core radius.

- All BCGs galaxies are born with an intrinsic GC population of $S_N \simeq 4$, similar to that of other elliptical galaxies. Galaxies with $S_N \gtrsim 4$ have gained an additional "excess" GC component. The number of excess GCs can be calculated using Equation 1,

$$N_{excess} = N_t - 4 \times 10^{-0.4(M_v+15)} \quad (4)$$

Values of $N_{excess}$ for each BCG are listed in Table 2.

- A galaxy may be surrounded by a halo of IGCs, the number of which will depend on its position in the host galaxy cluster.

Combining Equations 2 and 3, it is straightforward to estimate the projected local density of IGCs at any particular location in a cluster as

$$\Sigma_{IGC} \propto \frac{M_{cl}}{(1 + r^2/r_c^2)} \quad (5)$$

Accurate dynamical determinations of cluster masses from optical data are somewhat problematic due to the prevalence of substructure in many clusters of galaxies (see West 1994 for a review). A less ambiguous method is to use X-ray determinations of cluster masses. Under the assumption that the hot X-ray gas is in hydrostatic equilibrium in the cluster gravitational potential well, the X-ray temperature is expected to be linearly proportional to the total cluster mass ($T_X \propto M_{cl}$), which upon substitution into Equation 5 yields

$$\Sigma_{IGC} \propto \frac{T_X}{(1 + r^2/r_c^2)} \quad (6)$$

The cluster core radius, $r_c$ determines how centrally concentrated the IGC population is. Until recently it was thought that galaxy clusters possess core radii on the order of several hundred $h^{-1}$ kpc. However mounting observational evidence has shown that the mass distribution in the centers of clusters is much steeper than previously realized (e.g., Beers & Tonry 1986; Merrifield & Kent 1989; Gerbal et al. 1993). The most direct probe comes from observations of gravitational lensing, which have shown that the dark matter distribution is quite centrally concentrated, with typical core radii $r_c \simeq 20 - 50\,h^{-1}$ kpc (e.g., Kneib et al. 1993; Mellier, Fort & Kneib 1993; Miralda-Escudé 1995; Smail et al. 1995). As a first approximation, we shall adopt a core radius $r_c \simeq 30\,h^{-1}$ kpc for all clusters.

A natural consequence of this model is that BCGs, which usually reside at the dynamical centers of their parent clusters, are likely to be surrounded by extended halos of IGCs. Hence most of the excess GCs associated with high-$S_N$ galaxies may belong to a background population of intracluster globulars. Note, however, that location in a rich galaxy cluster is a necessary, but not sufficient, condition for producing high-$S_N$ galaxies. If a galaxy is significantly offset from the dynamical center of the cluster then it will not be surrounded by appreciable numbers of IGCs, since $\Sigma_{IGC}$ falls off rapidly with distance from the cluster center. Likewise, a BCG residing at the center of a poor cluster is unlikely to be a high-$S_N$ system simply because there are fewer IGCs in low mass galaxy clusters.

## 3. Results

Equation 6 can be used to estimate the local IGC density around each of the BGCs in Table 1. Three pieces of information are required: the X-ray temperature of the host galaxy cluster, its dynamical center and the position of the galaxy relative to this center.

X-ray temperatures for all clusters in Table 1 were obtained from the compilation by David et al. (1993), with the exception of the M87 and M49 subclusters, which were taken from Böhringer et al. (1994). Three of the poorest clusters in Table 1 do not have measured temperatures owing to their low X-ray luminosities (MKW 12, AWM 3, Abell 2666). In those cases, X-ray temperatures were estimated using the observed cluster velocity dispersion (from Beers et al. 1995 and Scoddegio et al. 1995) together with the well-known correlation between X-ray temperature and velocity dispersion (Lubin & Bahcall 1993).

Cluster centroids were also determined from X-ray observations, primarily the catalogue of Jones & Forman (1995), together with ROSAT observations of Virgo (Böhringer et al. 1994), Coma (Vikhlinin, For-man & Jones 1994) and Fornax (Rangarajan et al. 1995). The projected distance, $r$, of the BCG from the cluster center is listed in Table 2 along with the computed value of $\Sigma_{IGC}$. For those clusters with substructure, the distance of the BCG from the nearest subcluster centroid was used. For the three poor clusters without published X-ray maps, the BCG was assumed to reside at the cluster center.

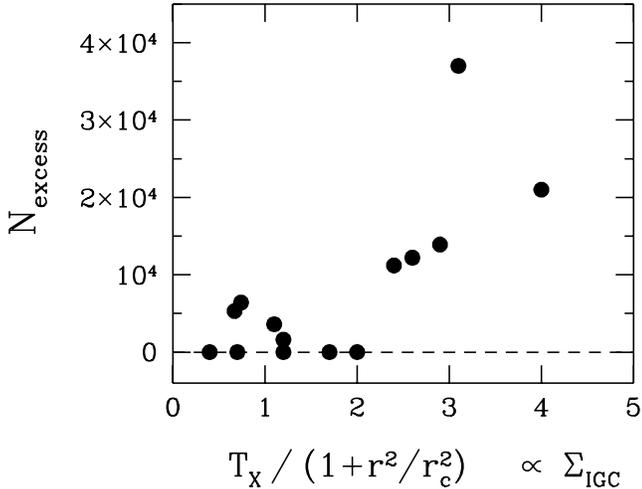

Figure 2. $N_{excess}$, the observed number of "excess" globular clusters associated with a BCG, versus the estimated local density of intracluster globular clusters, $\Sigma_{IGC}$.

A plot of $N_{excess}$ versus the estimated local IGC density, $\Sigma_{IGC}$, is shown in Figure 2. A correlation between these two quantities is clearly seen – those BCGs located in the densest cluster environments invariably have the largest excess GC populations. This correlation is all the more remarkable when one considers the large uncertainties in the observational data ($S_N$ values are typically uncertain by a factor of two) and the simplicity of the model employed here; the true correlation is presumably even stronger.

Closer examination of several galaxies serves to illustrate the important features of the IGC model. The galaxy with the largest measured specific frequency, UGC 9799 ($S_N = 20$), resides at the dynamical center of a rich galaxy cluster, Abell 2052. NGC 7768 ($S_N = 4$) also resides at the center of its parent cluster, however Abell 2666 is much poorer than Abell 2052, and hence would be expected to have few, if any, IGCs. The disparity between the specific frequencies of the two supergiant elliptical galaxies in the Coma cluster can now also be easily understood; NGC 4874 ($S_N = 14$) is located much closer to the cluster X-ray centroid than is NGC 4889 ($S_N = 7$), and thus should be surrounded by a greater density of IGCs.

## 4. The Origin of Intracluster Globular Clusters

What is the origin of the proposed population of IGCs?

One possibility is that they are tidally-stripped debris from galaxy collisions which has accumulated at the bottom of galaxy cluster potential wells (e.g., White 1987; Muzzio 1987). However, as van den Bergh (1984) has pointed out, a problem with this idea is that one would expect halo stars and GCs to be stripped in equal proportions during galaxy interactions, and hence accretion of such material by BCGs should not result in a net increase of $S_N$. Yet because of the mass-dependence of dynamical friction, it is possible that stripped GCs would settle much more readily toward the cluster center than stripped halo stars.

Alternatively, it is possible that IGCs might have formed *in situ* in the intracluster environments of galaxy clusters. In some GC formation scenarios (e.g., West 1993; Harris & Pudritz 1994), the efficiency of globular cluster formation is expected to depend sensitively on the local matter density. In the cores of galaxy clusters, the local density may have been sufficiently high to allow the birth of IGCs without the need for a parenting galaxy. It has also been suggested that IGCs might condense out of cooling flows (Fabian et al. 1984), although no correlation exists between $S_N$ and present-day cooling flow rate (Harris et al. 1995). Another speculative possibility might be that IGCs form during mergers of gas-rich subclusters of galaxies.

## 5. Conclusions

We have shown that a simple model based on the hypothesis of a population of intracluster globular clusters can explain the origin of the high-$S_N$ phenomenon, and the variation of GC populations among brightest cluster galaxies. High-$S_N$ galaxies did not form GCs more efficiently than other galaxies, but rather they have inherited an additional population of IGCs thanks to their fortuitous location at the dynamical centers of rich galaxy clusters.

Additional support for this idea comes from the large velocity dispersions of GC systems around M87 (Mould et al. 1987) and NGC 1399 (Grillmair et al. 1994), which suggest that these GCs are bound to the potential well of the galaxy cluster as a whole.

Any successful model should strive not only to explain existing observations, but also to make predictions which can be tested. In this spirit, we have selected a number of nearby ($z \leq 0.04$) BCGs whose GC populations have not yet been measured, and we attempt to predict in Table 3 which of these should be high-$S_N$ systems based on the X-ray temperatures of their parent clusters and the galaxy's location relative to the cluster centroid.

Finally, we note that intracluster globular clusters should be detectable with deep, high-resolution images of the cores of rich galaxy clusters, which provides the most stringent test of the IGC model proposed here. Such a population may also offer a valuable tool for extragalactic research by providing a new and independent tracer of the dark matter distribution in the cores of galaxy clusters.

M.J.W. was supported by the NSERC of Canada. C.J. and W.F. were supported by the Smithsonian Institution and the *AXAF* Science Center NASA Contract NAS8-39073.

TABLE 1
Globular Cluster Populations of Brightest Cluster Galaxies

| Galaxy | Cluster | $z$ | $M_V$ | $N_t$ | $S_N$ | $kT$(keV) | Reference |
|---|---|---|---|---|---|---|---|
| NGC 4486 (M87) | Virgo | 15 Mpc | -22.7 | 16000 | 14 | 2.4 | Harris 1986 |
| NGC 4472 (M49) | Virgo | 15 Mpc | -22.9 | 7400 | 5 | 1.2 | Harris 1986 |
| NGC 1399 | Fornax | 15 Mpc | -21.2 | 4800 | 16 | 1.1 | Bridges et al. 1991 |
| NGC 3311 | A1060 | 0.012 | -22.7 | 19000 | 15 | 3.9 | McLaughlin et al. 1995 |
| NGC 5629 | AWM3 | 0.015 | -21.7 | < 2000 | < 5 | $0.4^a$ | Bridges & Hanes 1994 |
| NGC 5424 | MKW12 | 0.019 | -21.6 | < 2000 | < 5 | $0.7^a$ | Bridges & Hanes 1994 |
| NGC 4073 | MKW4 | 0.020 | -23.1 | 7200 | 4 | 1.7 | Bridges & Hanes 1994 |
| NGC 3842 | A1367 | 0.021 | -23.2 | 14000 | 7 | 3.7 | Butterworth & Harris 1992 |
| NGC 4874 | A1656 | 0.023 | -22.7 | 17000 | 14 | 8.3 | Blakeslee & Tonry 1995 |
| NGC 4889 | A1656 | 0.023 | -23.2 | 13000 | 7 | 8.3 | Blakeslee & Tonry 1995 |
| NGC 7768 | A2666 | 0.028 | -22.9 | 6000 | 4 | $1.2^a$ | Harris et al. 1995 |
| NGC 6166 | A2199 | 0.031 | -23.6 | 10000 | 4 | 4.5 | Pritchet & Harris 1990 |
| UGC 9799 | A2052 | 0.035 | -23.4 | 46000 | 20 | 3.1 | Harris et al. 1995 |
| UGC 9958 | A2107 | 0.042 | -23.4 | 30000 | 13 | 4.2 | Harris et al. 1995 |

[a] Estimated X-ray temperature.

Table 2

Intracluster Globular Cluster Populations

| Galaxy | $N_{excess}$ | $r(h^{-1}$ kpc) | $\Sigma_{IGC}$ |
|---|---|---|---|
| NGC 4486 (M87) | 11200 | 0 | 2.4 |
| NGC 4472 (M49) | 1600 | 0 | 1.2 |
| NGC 1399 | 3600 | 0 | 1.1 |
| NGC 3311 | 13900 | 18 | 2.9 |
| NGC 5629 | $0^a$ | 0 | 0.4 |
| NGC 5424 | $0^a$ | 0 | 0.7 |
| NGC 4073 | 0 | 0 | 1.7 |
| NGC 3842 | 6400 | 60 | 0.7 |
| NGC 4874 | 12200 | 45 | 2.6 |
| NGC 4889 | 5300 | 105 | 0.6 |
| NGC 7768 | 0 | 0 | 1.2 |
| NGC 6166 | 0 | 33 | 2.0 |
| UGC 9799 | 37000 | 0 | 3.1 |
| UGC 9958 | 21000 | 6 | 4.0 |

[a] $S_N = 4$ has been assumed.

Table 3
Predictions of the IGC Model

| High-$S_N$ Galaxies | Low-$S_N$ Galaxies |
|---|---|
| MCG-02-12-039 (Abell 496) | NGC 2329 (Abell 569) |
| CGCG 077-097 (Abell 2063) | NGC 2832 (Abell 779) |
| NGC 4696 (Abell 3526) | NGC 5400 (MKW 5) |
| NGC 5920 (MKW 3S) | NGC 5718 (MKW 8) |

---